# A Compact, Planar, Eight-Port Waveguide Power Divider/Combiner: The Cross Potent Superhybrid[*]

Christopher D. Nantista and Sami G. Tantawi[**]
Stanford Linear Accelerator Center, Stanford University, Stanford CA 94309

*Submitted to IEEE Microwave and Guided Wave Letters*

[*] Work supported by Department of Energy contract DE–AC03–76SF00515.
[**] Also with the Communications and Electronics Department, Cairo University, Giza, Egypt.

# A Compact, Planar, Eight-Port Waveguide Power Divider/Combiner: The Cross Potent Superhybrid

Christopher D. Nantista and Sami G. Tantawi, *Member, IEEE*

*Abstract*-- In this letter, we present a novel four-way divider/combiner in rectangular waveguide. The design is completely two-dimensional in the h-plane, with eight-fold mirror symmetry, and is based on a recent four-port hybrid design [6]. In combining mode, it can function as a phased array with four inputs and four outputs. The planar nature of this design provides advantages, such as the freedom to increase the waveguide height beyond the over-moding limit in order to reduce field strengths. Along with its open geometry, this makes it ideal for high-power applications where rf break down is a concern. Design criteria, field-solver simulation results, and prototype measurements are presented.

*Index Terms*-- waveguide, hybrid, phased array, divider/combiner

## I. INTRODUCTION

Designs for a next-generation 0.5–1 TeV center-of-mass electron/positron collider [1],[2] call for power combined from each group of several 11.424 GHz 75 MW klystrons to power spacially separated sets of high-gradient accelerator structures of the main linacs through a Delay Line Distribution System (DLDS) [3],[4]. The rf is directed appropriately through source phasing during different time bins, yielding a pulse compression effect, in which each accelerator feed sees a peak power several times higher than that of a single source for a fraction of the klystron pulse width.

The basic requirement at the heart of this scheme is a phased array capable of directing pulsed rf from four inputs to four outputs (three delay lines and a local feed) at peak power levels reaching 600 MW. This can be accomplished with a set of hybrids and a pattern of discrete phase changes in the source drive signals. In prototype rf systems employing WR90 magic T's under high-vacuum for combining and directing power, rf breakdown proved to be a problem at power levels much exceeding 200 MW [5]. Autopsies revealed the problem to be between the matching element and the mouth of the e-bend.

This work was supported by the Department of Energy under contract DE-AC03-76SF00515.
The authors are with the Stanford Linear Accelerator Center, Stanford, CA 94309.
Sami G. Tantawi is also with the Communications and Electronics Department, Cairo University, Giza, Egypt.

This motivated the design of an alternative to the magic T, a novel planar hybrid capable of reliably handling hundreds of megawatts of peak power at X-band. Small apertures, slots, matching posts, and field-enhancing e-bends were avoided to reduce the risk of rf breakdown. This new hybrid, reported on in [6], consists of four rectangular waveguide ports, operated in the $TE_{10}$ mode, connected through two h-plane T-junctions, yielding an "H" geometry. The connecting waveguide carries two modes.

A further development led to a component, based on the same design technique, which provides the desired function of a set of four hybrids in a single compact device. This eight-port combiner/splitter has been named the "cross potent superhybrid", after the heraldic symbol its geometry recalls. Each port is matched, isolated from three other ports, and coupled equally to the four remaining ports. As the above-mentioned hybrid, or "magic H" serves as a sort of stepping stone to the cross potent superhybrid, we shall recount here the theory, design method and advantages of the former device.

## II. MAGIC H HYBRID

The magic H is a quadrature hybrid (i.e. the coupled port fields are 90° out of phase with each other) with directly opposite port pairs isolated. It can be viewed as a variation of the branchline coupler with the two connecting transmission lines collapsed into one waveguide utilizing two modes. A closer comparison might be made with the Riblet short-slot coupler [7], which employs the same coupling mechanism, although the "H" geometry provides separated ports and no mismatch due to wall thickness.

A key feature of this hybrid is its planar design. Matching features maintain the translational symmetry in the direction of the electric fields, so that the fields terminate only on the flat top and bottom surfaces. This H-plane symmetry allows the use of over-moded rectangular waveguide, in which the height has been increased to reduce field amplitudes, without affecting the scattering matrix. Theoretically, the height of the device can thus be arbitrarily increased to accommodate higher power.

The geometry of the magic H hybrid is illustrated in Figure 1. The connecting waveguide is wide enough to accommodate both $TE_{10}$ and $TE_{20}$ mode propagation. The fields of a single port excite these modes with equal amplitudes and a relative phase such that their fields add constructively on the side nearest the input port and destructively on the other side. As their guide wavenumbers are different, their relative phase changes as they propagate along the guide. A net phase slip of $\pi$ radians would cause the $TE_{10}$ wave to enhance the opposite lobe of the $TE_{20}$ wave, sending the power out the farthest port. To get a 3-dB split, therefore, the total phase lengths for these two modes must differ by an odd multiple of $\pi/2$.

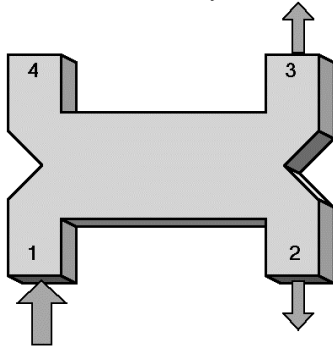

**Figure 1.** H-planar geometry of the magic H hybrid. Power-flow arrows illustrate splitting function (reverse for combining).

The T-junctions are matched by shaping the walls with blunt, triangular protrusions at the symmetry plane. The result is essentially a side-wall coupler with the common wall removed and two back-to-back mitred 90° bends at either end. The connecting guide must be narrower than twice the standard guide width in order to keep the $TE_{30}$ mode cut off. With the ports constrained to half the width of the connecting guide, standard reference curves or tables can be used to determine the dimension of the 45° mitres. The $TE_{20}$ (odd) mode is matched independently at each junction, as it sees an effective mitred bend. The $TE_{10}$ (even) mode is matched for discrete lengths of the connecting guide, at which the mismatches of the two junctions cancel. The width of the connecting guide is adjusted until a reasonable matching length also yields the required $\pi/2$ phase length difference through the hybrid for the two modes. The code HFSS [8] was used extensively in the design process to calculate scattering matrices.

### III. CROSS POTENT SUPERHYBRID

Often, a microwave circuit containing a configuration of multiple hybrids is required for a particular application. As mentioned above, the Delay Line Distribution System [3], conceived as a means of efficiently delivering high-power rf to the particle accelerator structures of a next-generation linear collider, requires power from four sources to be combined and delivered sequentially (by means of phase shifts) to four output waveguides. This can be done with four hybrids, two stages of two. One output port of each of the first two hybrids is connected to one input port of each of the second two, so that the combined circuit has four inputs and four outputs. The two sources feeding each first-stage hybrid are phased so as to combine in one or the other of its output ports (canceling in the fourth port). These combined waves are directed to the same second-stage hybrid during a given time bin. The relative phase between the source pairs provides another degree of freedom, so that the power arriving at the second-stage hybrid can be combined and directed to one or the other of its output ports. This arrangement is illustrated schematically in Figure 2.

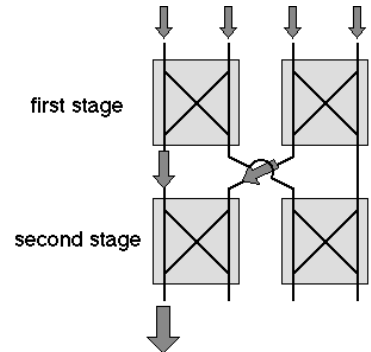

**Figure 2.** Schematic of hybrid arrangement for combining and directing four input signals to any of four outputs.

The same functionality can be achieved in a single eight-port device based on the magic H. Imagine placing two such such hybrids side-by-side and removing the common wall, leaving only the diamond shaped post formed by the two triangular mitre protrusions. The two combined ports formed are indistinguishable from the interior connecting guides. That is, the junction around the diamond-shaped post is four-fold symmetric. Adding mitred splits to these double-width ports at the proper distance effectively completes a second pair of hybrids. The resulting device, resembling a cross with cross beams at the extremities has been christened the cross potent superhybrid. The geometry is shown in Figure 3.

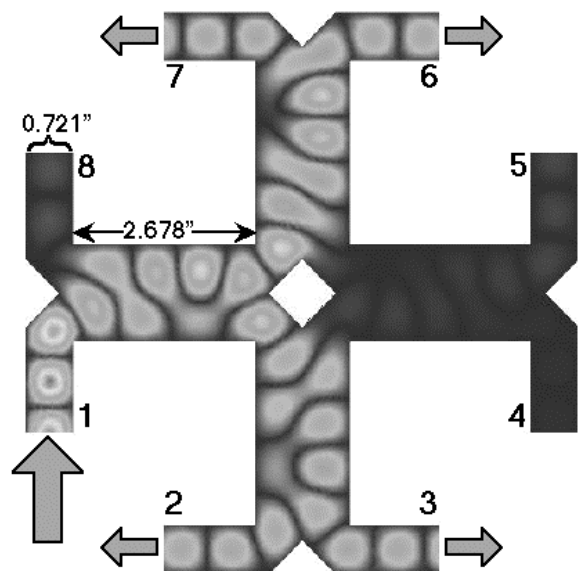

**Figure 3.** H-planar geometry of an eight-port cross potent superhybrid. Power-flow arrows and an HP HFSS electric field plot illustrate four-way splitting function with power fed into port 1 (reverse for combining).

Note that the four ports at the ends of the horizontal piece couple to the four ports at the ends of the vertical piece. This corresponds to a slightly different orientation of the hybrids in Figure 2 (albeit with the same connections) which circumvents the need to have one port connection cross over another, thus allowing planar symmetry to be maintained. The four magic H's which can be discerned in the cross potent superhybrid coalesce in such a way that the connected ports disappear entirely into the interiors of the neighboring hybrids, elegantly yielding a device which is more compact than the sum of its parts.

The HP HFSS [9] electric field plot in Figure 3 illustrates the power direction for a wave entering the lower left port. The ideal scattering matrix, with properly chosen reference planes, can be written as follows:

$$\mathbf{S}_{cp} = \frac{1}{2} \begin{bmatrix} 0 & 1 & -i & 0 & 0 & -1 & -i & 0 \\ 1 & 0 & 0 & -i & -1 & 0 & 0 & -i \\ -i & 0 & 0 & 1 & -i & 0 & 0 & -1 \\ 0 & -i & 1 & 0 & 0 & -i & -1 & 0 \\ 0 & -1 & -i & 0 & 0 & 1 & -i & 0 \\ -1 & 0 & 0 & -i & 1 & 0 & 0 & -i \\ -i & 0 & 0 & -1 & -i & 0 & 0 & 1 \\ 0 & -i & -1 & 0 & 0 & -i & 1 & 0 \end{bmatrix}$$

$$= \frac{1}{\sqrt{2}} \begin{bmatrix} \mathbf{S}_H & -\mathbf{S}_H{}^* \\ -\mathbf{S}_H{}^* & \mathbf{S}_H \end{bmatrix}.$$

Here $\mathbf{S}_H$ is the scattering matrix of the simple hybrid.

A cold-test model of the cross potent superhybrid has been built. A waveguide height of 0.400" was used, and 90° curved bends were incorporated to simultaneously turn the ports outward and taper from the 0.721" port width to 0.900". This allowed for easy testing with WR90 connectors and loads. Figure 4 shows the network analyzer results. $S_{71}$, being by symmetry and reciprocity identical to $S_{31}$, obscures the latter. All isolations are better than −38.8 dB at the design frequency and remain below −20 dB over a bandwidth of ~200 MHz. All couplings are within 0.06 dB of an average of −6.07 dB, adjusted for the added bends. Insertion loss is thus calculated to be ~0.05 dB. This device inherits all the advantages of the planar magic H hybrid, making it suitable for over-height fabrication for very high power applications.

## IV. CONCLUSIONS

In response to the problem of rf breakdown in multi-hundred-megawatt X-band rf systems being developed for a next generation linear collider, we have conceived and designed a rectangular waveguide, eight-port superhybrid, capable of serving as a 4x4 phased array, with a relatively open interior and a completely two-dimensional geometry. The latter feature makes its circuit properties independent of height, allowing for fabrication in over-height waveguide to minimize field strengths.

Measurements of a prototype were very satisfactory and in good agreement with simulation. Field plots from the latter suggest that the maximum field in our X-band design can be limited to 50 MV/m at 600 MW with a waveguide height of 1.35", whereas the same power in a magic T would yield surface fields exceeding 100 MV/m. A prototype of a similar four-port hybrid performed quite well [5] at high power. The simplicity of both these devices makes them attractive for low power use as well.

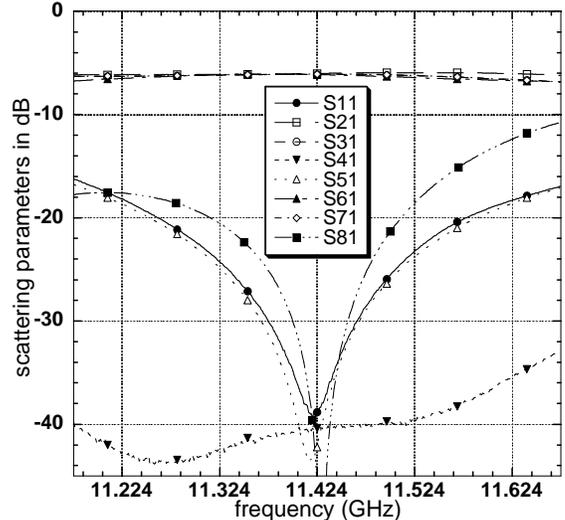

**Figure 4.** Measured amplitudes, in dB, of the first column scattering matrix elements (and by symmetry all others) for the cross potent superhybrid over a frequency range of 500 MHz centered on 11.424 GHz.

## V. REFERENCES

[1] The NLC Design Group, Zeroth-Order Design Report for the Next Linear Collider, LBNL-PUB 5424, SLAC Report 474, and UCRL-ID 124161, May 1996.

[2] The JLC Design Group, JLC Design Study, KEK-REPORT-97-1, KEK, Tsukuba, Japan, April 1997.

[3] H. Mizuno and Y. Otake, "A New RF Power Distribution System for X Band Linac Equivalent to an RF Pulse Compression Scheme of Factor $2^N$," contributed to the 17th International Linac Conference (LINAC94), Tsukuba, Japan, August 21−26, 1994.

[4] S.G. Tantawi *et al.*, "A Multi-Moded RF Delay Line Distribution System for the Next Linear Collider," proceedings of the 8th Workshop on Advanced Accelerator Concepts, Baltimore, Maryland, July 5−11 1998.

[5] A.E. Vlieks *et al.*, "High Power RF Component Testing for the NLC," presented at the 19th International Linear Accelerator Conference (LINAC 98) Chicago, IL, August 23−28, 1998, SLAC-PUB-7938 (1998).

[6] C.D. Nantista, *et al.*, "Planar Waveguide Hybrids for Very High Power RF," presented at the 1999 Particle Accelerator Conference, New York, NY, March 29−April 2, 1999.

[7] Henry J. Riblet, "The Short-Slot Hybrid Junction," Proceedings of the I.R.E., February 1952, p. 180.

[8] High Frequency Structure Simulator, Version A.04.01, copyright 1984−1995 Ansoft Corp., copyright 1990−1995 Hewlett-Packard Co.

[9] HP High Frequency Structure Simulator, Version 5.3, copyright 1996−1998 Hewlett-Packard Co.